\documentclass{aastex}
\usepackage{graphics}
           
\begin{document}
 
\title{ANOMALOUS ABSORPTION IN CYCLIC C$_3$H RADICAL}
\author{Suresh Chandra\altaffilmark{1} and S.A. Shinde }
\affil{School of Physical Sciences, S.R.T.M. University, \\
Nanded 431 606, India}
\email{sch@iuca.ernet.in}
\altaffiltext{1}{Visiting Associate, Inter-University Centre for Astronomy 
\& Astrophysics, Pune \mbox{411 007}, India} 
\begin{abstract}
Yamamoto et al. (1987) reported the first detection of $c$-C$_3$H radical 
in \mbox{TMC-1} through its transition $2_{1 2} \rightarrow 1_{1 1}$ 
at 91.5 GHz. Mangum and Wootten (1990) detected $c$-C$_3$H through the 
transition $1_{1 0} \rightarrow 1_{1 1}$ at 14.8 GHz in 12 additional 
galactic objects. The column density of $c$-C$_3$H in the objects was 
estimated to be about one order of magnitude lower than that of the 
C$_3$H$_2$ which is ubiquitous in the galactic objects. The most probable 
production mechanism of both the C$_3$H and C$_3$H$_2$ in dark clouds
is a common dissociation reaction of C$_3$H$_3^+$ ion (Adams \& Smith, 1987).
Although the $c$-C$_3$H is 0.8 eV less stable than its isomer $l$-C$_3$H, 
finding of comparable column densities of both the isomers in \mbox{TMC-1}
supports the idea of comparable formation of both the $c$-C$_3$H and $l$-C$_3$H
in the cosmic objects.  Existence of a  metaisomer in interstellar condition 
is a well known phenomenon in astronomy. 

We propose that $c$-C$_3$H may be identified through the 
transition $3_{3 1} \rightarrow 3_{3 0}$ at 3.4 GHz in absorption against 
the cosmic 2.7 K background in dense cosmic objects when no strong source 
is present in the background. When there is some strong source in the 
background of the object, peak of the absorption line decreases with the 
increase 
of the strength of the background source. However, at low densities, the 
intensity is found to increase. Hence, in low density regions, a background
source can help in detection of the line. This absorption line may play an important
role for identification of $c$-C$_3$H in a large number of cosmic objects. Similar absorption features are found 
for $c$-C$_3$D radical also.
\end{abstract}

\section{\sc introduction}
Discovery of hydrocarbon ring molecules in interstellar medium has provided 
an important tool to investigate the medium. Cyclopropenylidene (cyclic 
C$_3$H$_2$, denoted as $c$-C$_3$H$_2$) was the first hydrocarbon ring molecule 
discovered in astronomical objects by Thaddeus et al. (1985b). This molecule 
is ubiquitous in the galactic objects (Matthews and Irvine, 1985), and
except in one case, its line $2_{1 1} \rightarrow 2_{2 0}$  at 21.59 GHz 
has always been found in absorption against the cosmic 2.7 K background
(also called  the cosmic microwave background, denoted as CMB)
(Madden et al., 1989; Chandra \& Kegel, 2001). Cox et al. (1987), however, 
reported this line in emission in the Planetary Nebula NGC 7027. The second 
hydrocarbon ring molecule, cyclopropynylidyne (cyclic C$_3$H, denoted as 
$c$-C$_3$H) was first detected by Yamamoto et al. (1987) in TMC-1  
through its transition $2_{1 2} \rightarrow 1_{1 1}$ at 91.5 GHz. 
Mangum and 
Wootten (1990) detected $c$-C$_3$H through the transition $1_{1 0} 
\rightarrow 1_{1 1}$ in 12 additional galactic objects. The $l$-C$_3$H (linear
C$_3$H) also
has been identified in astronomical objects (Thaddeus et al., 1985a). The most 
probable production mechanism of C$_3$H and C$_3$H$_2$ in dark clouds 
is a common  dissociation reaction of C$_3$H$_3^+$ ion (Adams \& Smith, 1987):
\vspace{2mm}

\begin{tabular}{ll}
C$_3$H$_3^+$ + e &  
 $\rightarrow$ C$_3$H + H + H \hspace{5mm} or \hspace{5mm}
 C$_3$H + H$_2$ \\
& $\rightarrow$ C$_3$H$_2$ + H \\
\end{tabular}

\vspace{2mm}

\noindent
Although $c$-C$_3$H is 0.8 eV less stable than $l$-C$_3$H, the column
density of $c$-C$_3$H is found almost comparable to that of $l$-C$_3$H
in TMC-1. Such a metastable isomer can exit in the interstellar
condition. For example, the HNC, the isomer of HCN, is detected in various
molecular clouds. 
In the preset $Letter$, we propose to identify $c$-C$_3$H in cosmic objects
through $3_{3 1} \rightarrow 3_{3 0}$ transition in absorption against the 
CMB.

\section{\sc molecular data \& calculations}

The $c$-C$_3$H is an $a$-type asymmetric top molecule with electric dipole
moment $\mu$ = 2.4 Debye. Rotational levels in an asymmetric top molecule 
are specified as $J_{k_a, k_c}$, where $J$ is the rotational quantum number,
and $k_a$ and $k_c$ the projections of $J$ on the axis of symmetry in case
of prolate and oblate symmetric tops, respectively. In $a$-type asymmetric 
top molecule, rotational radiative transitions are governed by the 
selection rules:

\vspace{2mm}

\begin{tabular}{rl}
$J :$ & $\Delta J = 0 \pm 1$ \\
$k_a, k_c:$ & even, odd $\longleftrightarrow$ even, even \\
         :  & odd, even $\longleftrightarrow$ odd, odd \\
\end{tabular}

\vspace{2mm}

\noindent
Owing to the nuclear symmetry in $c$-C$_3$H, rotational levels with even
value of $k_a$ are missing in its spectra. As the temperature in dark 
molecular clouds 
is not large, in the present investigation we accounted for the rotational 
levels up to 77 cm$^{-1}$ for $c$-C$_3$H (Table 1). The Einstein 
$A$-coefficient for rotational transitions between the levels are calculated 
following the method discussed by Chandra \& Sahu (1993) and Chandra \& 
Rashmi (1998). These 37 rotational levels are connected through 121 
radiative transitions. For the calculations, the required molecular and 
distortional constants are taken from Yamamoto and Saito (1994). We have not,
however, accounted for fine structure and hyperfine structure splittings of 
the levels. 

NLTE occupation numbers of energy levels are calculated in an 
$on$-$the$-$spot$ approximation discussed by Rausch et al. (1996), where 
the external radiation field, impinging on a volume element generating 
the lines, is the CMB  only. Besides the radiative transition probabilities 
for radiatively allowed transitions 
between the rotational levels, data required for the present 
investigation are the rate coefficients for collisional transitions between
the levels due to collisions with H$_2$ molecules. Collisional rate
coefficients are not available in literature. Therefore, the rate 
coefficients for downward transitions  $J'_{ k'_a  k'_c} \rightarrow 
J_{ k_a k_c}$ at a kinetic temperature $T$ are taken as (Sharma \& 
Chandra, 2001)
\begin{eqnarray}
C(J'_{ k'_a  k'_c} \rightarrow J_{k_a k_c}) = 1 \mbox{x} 10^{-11} \sqrt{T/30} /
(2 J'+1) \nonumber
\end{eqnarray}
\noindent
For upward collisional rate coefficients, we accounted for the fact that
downward and upward collisional rate coefficients are related through the
detailed equilibrium (Chandra \& Kegel, 2000).

\section{\sc results \& discussion}
In order to include a large number of cosmic objects where $c$-C$_3$H
may be found, numerical calculations are carried out for wide ranges of 
physical parametres. The molecular hydrogen density $n_{H_2}$ has been 
varied over the range from 10$^3$ cm$^{-3}$ to 10$^7$ cm$^{-3}$, and
calculations are performed for three kinetic temperatures 10, 20 and 30 K. 
In the calculations, free parametres are the molecular hydrogen density 
$n_{H_2}$, and $\gamma \equiv n_{mol}/(dv_r/dr)$, where $n_{mol}$ is 
density of the $c$-C$_3$H
molecule, and $dv_r/dr$ the velocity gradient. As we have used scaled values
of collisional rates, our results are qualitative in nature.

\subsection{\sc anomalous absorption in $c$-C$_3$H radical}
Out of a number of lines of $c$-C$_3$H found in absorption against the CMB, 
the transition $3_{3 1} \rightarrow 3_{3 0}$ at 3.4 GHz has 
shown reasonably good absorption phenomenon. The Einstein $A$-coefficient
for this line is 1.9 x 10$^{-9}$ s$^{-1}$, and the radiative lifetimes of
the upper and lower levels $3_{3 0}$ and $3_{3 1}$ are 2.3 x 10$^4$ s
and 4.4 x 10$^4$ s, respectively. Observation of a spectral line 
in absorption against the CMB is an unusual phenomenon. The intensity, 
$I_\nu$, of a line generated in an interstellar cloud, with homogeneous 
excitation conditions, is given by
\begin{eqnarray}
I_\nu - I_{\nu,bg} = (S_\nu - I_{\nu,bg}) (1 - e^{-\tau_\nu})\nonumber
\end{eqnarray}

 \noindent
where $I_{\nu,bg}$ is  the intensity of the continuum against which the 
line is observed, $\tau_\nu$  the optical depth of the line, and $S_\nu$ 
the source function. For positive optical depth, observation of an 
interstellar line in absorption against the CMB, obviously, implies the 
excitation temperature of the line, $T_{ex}$, to be less than the CMB 
temperature \mbox{2.7 K}, but positive. It requires rather peculiar 
conditions in the molecule generating the line.

Variation of line-intensity against the CMB in the units of the Planck's 
function at the kinetic temperature of $T$(K), i.e., $(I_\nu - 
I_{\nu, bg})/B_\nu(T)$, for the line $3_{3 1} \rightarrow 3_{3 0}$ 
for three kinetic temperatures 10, 20 and 30 K (written on the top of the
column) is shown in \mbox{Figure 1} 
(first row) for 
$\gamma = $ 10$^{-6}$ and 10$^{-5}$ cm$^{-3}$ (km/s)$^{-1}$ pc. The 
line $3_{3 1} \rightarrow 3_{3 0}$ shows the 
absorption against the CMB. The excitation temperature of the line 
reduces up to 0.27 K which 10\% of the CMB temperature,
and its absorption intensity is comparable to the absorption intensity of the
line $2_{1 1} \rightarrow 2_{2 0}$ of $c$-C$_3$H$_2$ calculated by Chandra \& 
Kegel (2001). It supports the idea of detection of the line 
$3_{3 1} \rightarrow 3_{3 0}$ of $c$-C$_3$H in absorption against the CMB.

\subsection{\sc effect of a background source}
In the above discussion, we considered the external radiation field,
impinging on a volume element generating the lines, to be the CMB
only. In order to see the effect of a background source, we repeated 
the calculations for the background intensity
\begin{eqnarray}
I_{\nu, bg} = (1-f) B_\nu(2.7 K) + f B_\nu(2000 K) \nonumber
\end{eqnarray}

\noindent
$f$ is the dilution factor, representing the distance of the background
source from the cosmic object, and the effect of the atmosphere between the 
source and the object. Here, the temperature of the background source is assumed
to be 2000 K. For other values of the temperature of the source, the effect
can be scaled down with the help of the dilution factor $f$. The case
$f = 0$ corresponds to the CMB only. With the increase of the value of $f$ 
the peak of the absorption line decreases. However,
at low densities, the intensity increases. Hence, in the low density regions,
a background source can help in detection of the line. \mbox{Figure 1} shows 
the absorption intensities for $f$ = 0, 10$^{-4}$ and 10$^{-3}$. When $f$ is
large than 10$^{-3}$, the absorption feature disappears.

\subsection{\sc anomalous absorption in $c$-C$_3$D radical}
We repeated the calculations for $c$-C$_3$D radical, where we accounted 
for the rotational levels given in Table 1. Note that the sequence of
some higher levels of $c$-C$_3$D is changed relative to those of the 
$c$-C$_3$H. This molecule also is an $a$-type asymmetric top molecule with 
electric dipole moment $\mu$ = 2.4 Debye. The required molecular and 
distortional constants are taken from Yamamoto and Saito (1994). Here, 
also the line $3_{3 1} \rightarrow 3_{3 0}$ at 0.9 GHz is found in 
absorption against the CMB. Variation of line-intensity against the CMB 
in the units of Planck's function at the kinetic temperature of $T$(K), i.e., 
$(I_\nu - I_{\nu, bg})/B_\nu(T)$, for the line for three kinetic temperatures 
10, 20 and 30 K and the dilution factor $f$ = 0, 10$^{-4}$ and 10$^{-3}$ 
is shown in 
\mbox{Figure 2} for $\gamma =$ 10$^{-6}$ and 10$^{-5}$ cm$^{-3}$ (km/s)$^{-1}$ pc. 
The line $3_{3 1} \rightarrow 3_{3 0}$ of $c$-C$_3$D is found to be weaker
than that of the $c$-C$_3$DH. However, the features of the two are very 
similar.

\acknowledgments
We re grateful to Prof. J.V. Narlikar, Prof. S.A. Suryawanshi, and Prof.
Dr. W.H.  Kegel for encouragement.

\clearpage

\begin{figure}
\includegraphics[width=15cm]{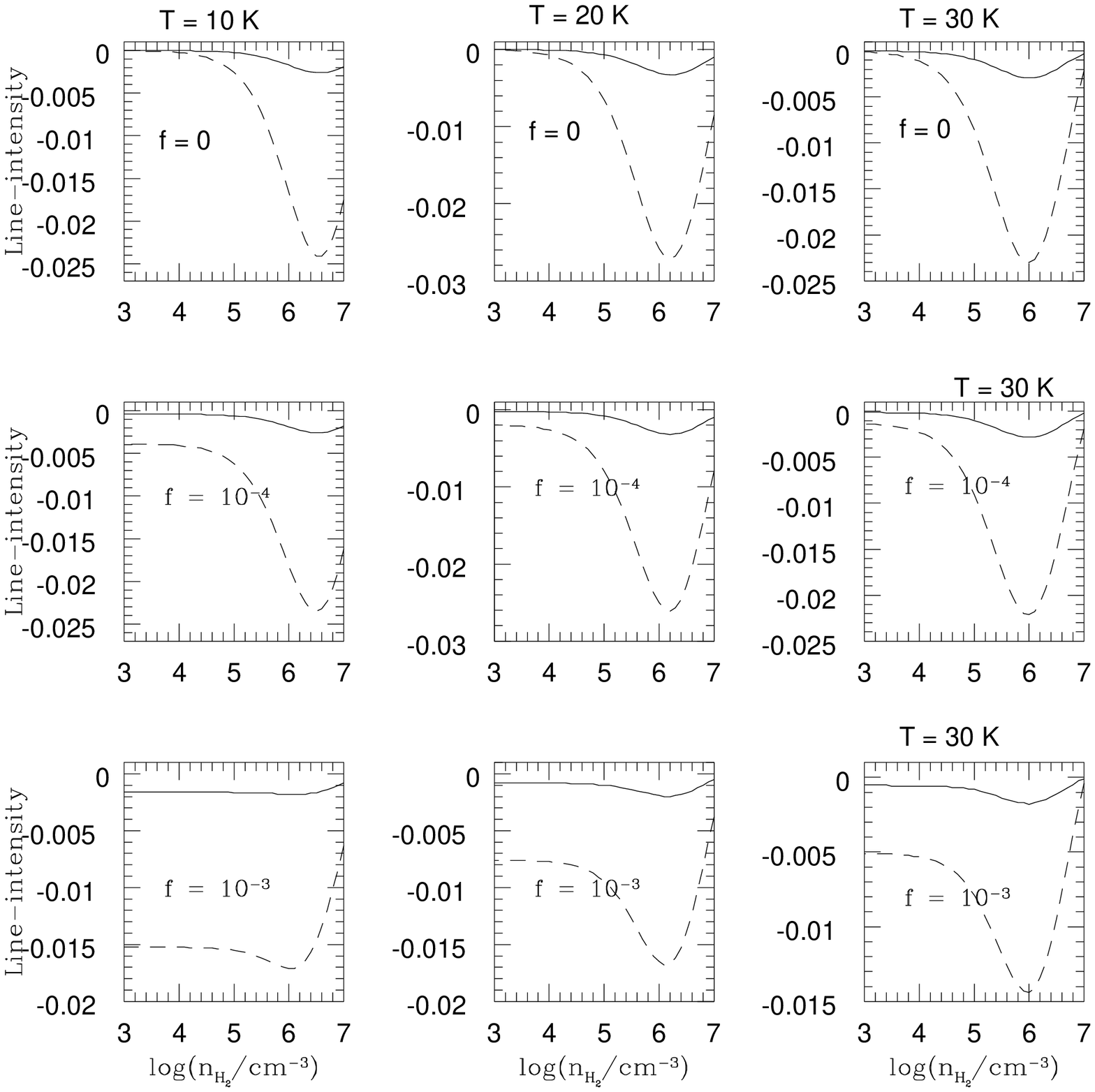}
\caption{Variation of line-intensity against the CMB in the units of 
Planck's function at the kinetic temperature of $T$(K), i.e., $(I_\nu 
- I_{\nu, bg})/B_\nu(T)$, for the transition $3_{3 1} \rightarrow 3_{3 0}$ 
of $c$-C$_3$H radical for three kinetic temperatures 10 20, and 30 K 
(written at the top of each column) with dilution factor $f$ written there.
Negative value of line-intensity shows absorption against the CMB. Solid 
line is for $\gamma =$ 10$^{-6}$ cm$^{-3}$ (km/s)$^{-1}$ pc and dotted 
line for $\gamma =$ 10$^{-5}$ cm$^{-3}$ (km/s)$^{-1}$ pc}
\end {figure}

\clearpage

\begin{figure}
\includegraphics[width=15cm]{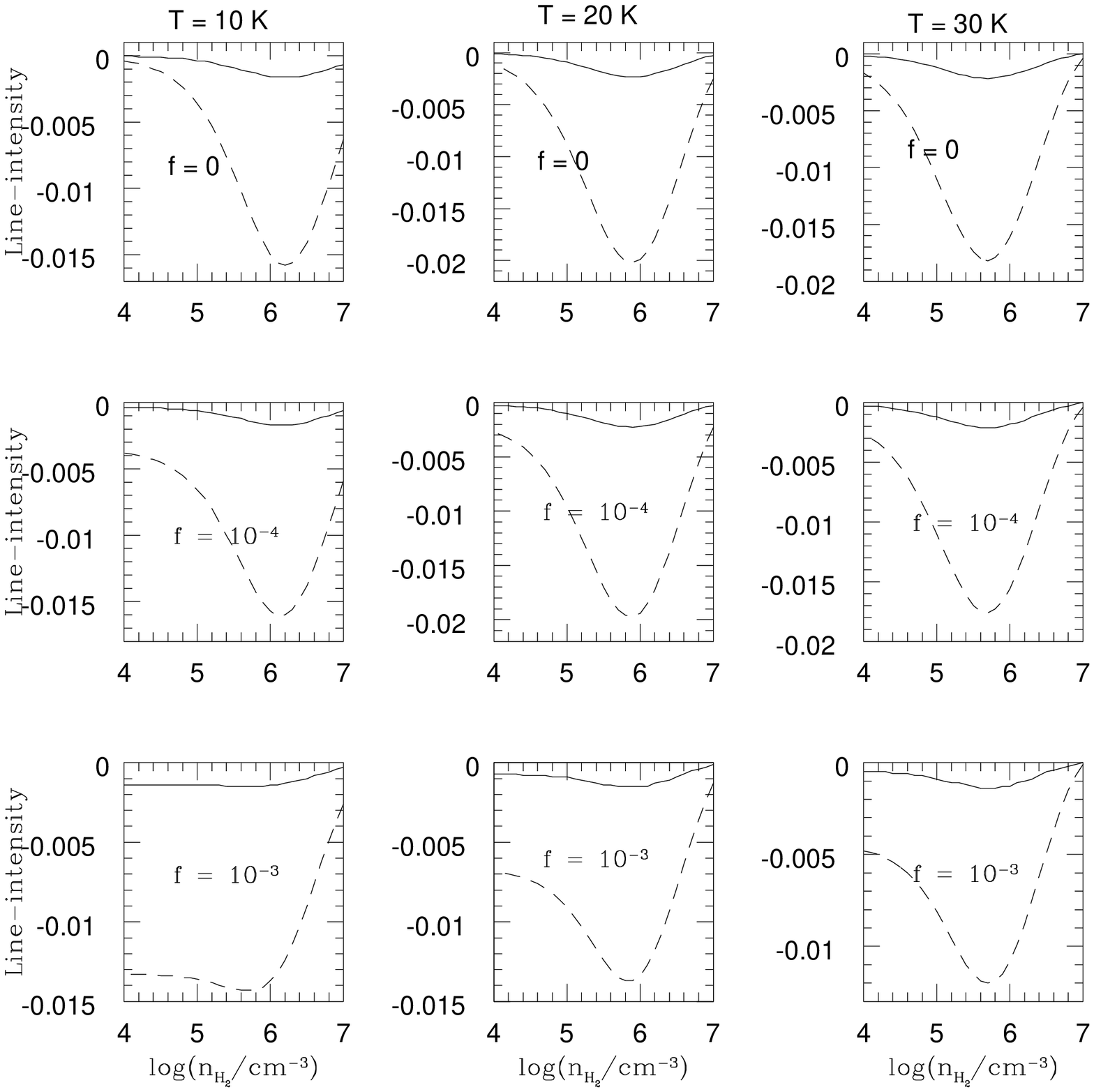}
\caption{Variation of line-intensity against the CMB in the units of 
Planck's function at the kinetic temperature of $T$(K), i.e., $(I_\nu 
- I_{\nu, bg})/B_\nu(T)$, for the transition $3_{3 1} \rightarrow 
3_{3 0}$ of $c$-C$_3$D radical for three kinetic temperatures 10 20 
and 30 K (written at the top of each column) with dilution factor $f$ 
written there.  Negative value of line-intensity shows absorption 
against the CMB. Solid line is for $\gamma =$ 10$^{-6}$ cm$^{-3}$
(km/s)$^{-1}$ pc and dotted line for $\gamma =$ 10$^{-5}$ cm$^{-3}$
(km/s)$^{-1}$ pc}
\end {figure}

\begin{table}
\begin{center}
\begin{tabular}{|rrr|r|r|rrr|r|r|}
\multicolumn{10}{l}{Table 1. Rotational energy levels of $c$-C$_3$H \&
$c$-C$_3$D radicals}\\
\hline
$J$ & $k_a$ & $k_c$ & \multicolumn{2}{c|}{$E(\mbox{cm}^{-1})$} &$J$ & $k_a$ & 
$k_c$ & \multicolumn{2}{c|}{$E(\mbox{cm}^{-1})$}\\
\cline{4-5} \cline{9-10}
&&& $c$-C$_3$H & $c$-C$_3$D &&&& $c$-C$_3$H & $c$-C$_3$D \\
\hline
  1 &   1 &   1 &   2.126  &   2.055  & 
  5 &   5 &   1 &  41.864  &  40.995 \\
  1 &   1 &   0 &   2.620  &   2.416  & 
  5 &   5 &   0 &  41.876  &  40.996 \\
  2 &   1 &   2 &   5.180  &   4.695  & 
  6 &   3 &   4 &  43.515  &  38.776 \\
  2 &   1 &   1 &   6.664  &   5.779  & 
  6 &   3 &   3 &  46.751  &  40.312 \\
  3 &   1 &   3 &   9.620  &   8.586  & 
  7 &   1 &   6 &  48.690  &  43.154 \\
  3 &   1 &   2 &  12.472  &  10.724  & 
  8 &   1 &   8 &  51.243  &  45.588 \\
  4 &   1 &   4 &  15.376  &  13.678  & 
  6 &   5 &   2 &  53.262  &  50.327 \\
  3 &   3 &   1 &  16.174  &  15.684  & 
  6 &   5 &   1 &  53.378  &  50.338 \\
  3 &   3 &   0 &  16.289  &  15.715  & 
  7 &   3 &   5 &  55.653  &  49.337 \\
  4 &   1 &   3 &  19.696  &  17.096  & 
  7 &   3 &   4 &  60.619  &  52.134 \\
  5 &   1 &   5 &  22.421  &  19.936  & 
  8 &   1 &   7 &  60.874  &  54.025 \\
  4 &   3 &   2 &  23.625  &  21.876  & 
  9 &   1 &   9 &  63.410  &  56.420 \\
  4 &   3 &   1 &  24.249  &  22.071  & 
  7 &   5 &   3 &  66.637  &  61.287 \\
  5 &   1 &   4 &  28.128  &  24.694  & 
  7 &   5 &   2 &  67.177  &  61.348 \\
  6 &   1 &   6 &  30.748  &  27.345  & 
  8 &   3 &   6 &  69.120  &  61.194 \\
  5 &   3 &   3 &  32.792  &  29.594  & 
  9 &   1 &   8 &  74.336  &  66.020 \\
  5 &   3 &   2 &  34.490  &  30.256  & 
  8 &   3 &   5 &  75.712  &  65.522 \\
  6 &   1 &   5 &  37.780  &  33.385  & 
 10 &   1 &  10 &  76.856  &  68.391 \\
  7 &   1 &   7 &  40.356  &  35.896  & 
    &     &     &          &         \\
\hline
\end{tabular} 
\end{center}
\end{table}

\end{document}